\documentclass[
 amsmath,amssymb,
 notitlepage,
superscriptaddress,
nofootinbib ]
{revtex4-2}

\usepackage{graphicx}
\usepackage{dcolumn}
\usepackage{bm}
\usepackage{braket}

\usepackage{xcolor}

\begin{document}

\title{Anti-Correlated Photons from Classical Electromagnetism}

\author{Ken Wharton}
\affiliation{Department of Physics and Astronomy, San Jos\'e State University, San Jos\'e, CA 95192-0106}
\author{Emily Adlam}
\affiliation{The University of Western Ontario, London, Canada}

\begin{abstract}
For any experiment with two entangled photons, some joint measurement outcomes can have zero probability for a precise choice of basis.  These perfect anti-correlations would seem to be a purely quantum phenomenon.  It is therefore surprising that these very anti-correlations are also evident when the input to the same experiment is analyzed via classical electromagnetic theory.  Demonstrating this quantum-classical connection for arbitrary two-photon states (and analyzing why it is successful) motivates alternative perspectives concerning entanglement, the path integral, and other topics in quantum foundations.

\end{abstract}

\maketitle

\section{INTRODUCTION}

Some quantum phenomena seem to have no classical explanation, but the list of inherently non-classical processes is much shorter than one might initially suppose.  While some may take superposition or interference to signify essentially quantum behavior, there are certainly entirely classical accounts of superposition and interference in the framework of electromagnetic (EM) fields.  Less well-known, there exist classical accounts of the essential features of quantum teleportation \cite{spekkens2007}, spin-1/2 systems \cite{ohanian1986}, and quantum tunneling \cite{zhu1986}.  And given that Einstein first introduced the idea of stimulated emission in a classical field context \cite{einstein1917}, it should be no surprise that there also exist detailed accounts of how classical field modes can become amplified in certain classical media \cite{fain1987, mertens2011}.  Furthermore, many authors have noted that any account of stimulated emission can be extended to \textit{spontaneous} emission by postulating an appropriate background field to act as a `seed' \cite{milonni1975,boyer1975}.  While such seeds are sometimes considered in context of quantum ``vacuum noise'' \cite{clerk2010}, in a classical analog such incoming background fields need not be quantum.  In this context, Boyer \cite{boyer1969} has noted that ``all formulations of classical e1ectromagnetism allow unspecified incoming radiation in the far past.'' 

Nevertheless, in the realm of phenomena related to quantum entanglement, classical analogs are famously hard to come by.  Classical accounts of entanglement are restricted by strong no-go theorems \cite{bell1995}, and cannot be achieved without relaxing criteria that most physicists would consider essentially ``classical'' in the first place \cite{wharton2020}.  So it would seem that only a quantum treatment of entanglement could correctly predict joint probabilities of distant measurements, including cases for which these joint probabilities equal zero.  A zero joint probability occurs when two distant possible outcomes are ``anti-correlated'', meaning that either outcome can occur on its own, but never both outcomes in the same run of the experiment.  This paper concerns anti-correlated photons, and demonstrates that such anti-correlations do indeed have a classical analog.  

At some level, this may not be surprising; after all, there are plenty of known classical anti-correlations (e.g., two players cannot both be dealt the same card from the same deck).  For a given polarization-entangled two-photon state, say $(\ket{HV}+\ket{VH})/\sqrt{2}$, one trivial model to explain the zero probability of an $\ket{VV}$ or $\ket{HH}$ outcome would be to suppose that classical EM waves were always sent to the two detectors such that one wave was horizontally-polarized and the other was vertically-polarized.  But such an \textit{ad hoc} model would not explain why such a preparation would correspond to this particular state, and it would not be generalizable to other evident anti-correlations from the very same state.  (The above quantum example also would exhibit anti-correlations if each photon were measured in the same diagonally-polarized basis, while the \textit{ad hoc} classical model would not.)  Furthermore, such a contrived preparation would not follow from classical EM theory, but would rather be something new.  The implication of the below results is that classical EM theory alone -- applied to the usual techniques by which entangled photons are generated -- can already explain the anti-correlations observed in this experiment, and many others besides.  Notably, the classical account of anti-correlations also extends to partially-entangled two-photon states, even if measured in the particular basis where three of the four joint outcomes are possible, and only one particular pair of outcomes is forbidden.

Of course, since no fully classical account can explain all entanglement correlations, any success at explaining the zero-probability joint outcomes must imply a failure to explain at least some of the non-zero probabilities.  It transpires that the classical analysis below will not be able to explain \textit{any} non-zero probabilities, or even explain why only two detectors fire in a 2-photon experiment.  Designing a different classical model to explain these features, such as the \textit{ad hoc} model in the previous paragraph, tends to be strictly basis-dependent, where changing the measurement settings causes the model to fail.  In contrast, the classical analysis of the preparation procedure yields robust results, even with a changing measurement basis.  

The essential argument in this paper is simple enough to present in its entirety in a single paragraph.  Entangled photons are commonly generated via spontaneous parametric down-conversion (PDC) from a pump laser.  As noted above, a classical account of spontaneous emission processes requires background input fields to seed the process. Leaving aside the thorny issue of the proper distribution -- or even the potential existence -- of the background input fields, one can still ask: According to classical EM, for any possible background fields, what are the allowed outputs from this experimental procedure?  The remarkable conclusion is that according to classical EM, there are some joint outcome patterns which can never extract energy from the pump laser, and these are precisely the outcome patterns for which quantum theory says should have zero probability. In other words, for any joint outcome which is assigned zero probability by quantum theory, there are no possible classical background seeds that will both result in the analogous\footnote{The precise analog will be clarified below, but there is evidently always a classical EM field corresponding to any set of measurable single-photon properties (frequency, energy, polarization, wavevector, etc.)} EM field output and also have a net amplification. Since there must be some amplification if energy is to be transferred from the pump laser, it follows that this joint outcome is also impossible in classical EM, for any conceivable input seeds. This result then motivates the conjecture that Joint Outcomes Require Classical Analogs (JORCA), or more specifically: \textit{If a pattern of output fields cannot result from amplifying some set of classical input fields, that pattern of outputs is impossible according to quantum theory}.

A single-photon version of this conjecture would seem to be obvious, following from the usual correspondence principle between classical intensity and quantum probability.  But it is far from obvious why this JORCA conjecture would apply to distant correlations, especially given that those correlations are known to violate Bell-inequalities and other no-go theorems.  It is hoped that this paper will stimulate a future answer to this foundational question, and in general  will further elucidate the distinction between classical and quantum systems.  It may also be of practical importance to quantum theory; if a complicated multi-photon entanglement scenario can be cast into classical terms (even if only concerning which joint outcomes are impossible) such a relationship might be of great practical benefit.  

Another implication of the JORCA conjecture would be the existence of a classical analog for any non-zero probability outcome in entanglement scenarios.   Specifically, in the measurements allowed by quantum theory, there always seems to be a pattern of classical input seed waves that would experience a net gain from a classical pump beam and then result in an output field pattern analogous to the measured photons.  To the extent that the classical background seeds were not detectable, such incoming fields would act as a ``hidden variable''.  However, it must be reinforced that this model makes no attempt to explain these non-zero-probability cases or assign probabilities to such outcomes.  (Indeed, any successful classical attempt to generate such probabilities would have to break one of the assumptions of Bell's Theorem.)  Still, the mere existence of these classical analogs is intriguing, and implications of this framework for non-zero probability outcomes will be explored in a future paper.

There exist earlier approaches aiming to explain Bell experiments based on hidden-variable models of PDC - for example, by Santos\cite{SANTOS199610}. However, these models make \textit{ad hoc} choices of hidden variables specially chosen to fit the observed results, and the zero-probability results were accounted for by imperfect detection efficiencies. By contrast, in this paper we develop a model which is based on well-recognised principles of classical EM, and demonstrates that the zero-probability outcomes of Bell experiments can be reproduced exactly.  There also exist approaches using quantum vacuum fields as seeds to explain PDC - for example, by  Heuer, Menzel, and Milonni\cite{Heuer_2015} - and approaches using the Wigner representation for PDC - for example, by Casado, Marshall and Santos\cite{Casado:97}. However, these models are quantum, whereas our model uses entirely classical EM, thus demonstrating that the quantum features of these previous approaches are not necessary to explain the zero-probability outcomes.  

The outline of this paper is as follows:  After detailing an evident classical analog of an entanglement-generating experiment in section II, the results will be successfully applied to the $(\ket{HV}+\ket{VH})/\sqrt{2}$ example in section III.A.  More sophisticated examples will be presented in the rest of section III, culminating in a classical analysis of a proof by Hardy \cite{hardy1993} which utilizes zero-probability outcomes to indicate quantum non-locality.  Section IV will provide generalizations to all PDC two-photon anti-correlations, as well as to most other techniques by which entangled photons can be created.  An analysis of how this classical account might happen to be successful is given in the final section, along with a number of future directions motivated by this intriguing result. 

\section{Classical Account of Entanglement Preparation}

If classical EM theory were exactly correct, it is not immediately obvious what it would predict for a typical experiment known to actually generate entangled photons.  This section will outline what would be predicted from such an experiment, given only classical EM theory and classical (but unknown) input seeds in the form of classical EM waves.  The detectors will also be treated classically, in that they will serve as a perfectly informative measurement of what fields are present at those locations.  (This is not a semiclassical analysis of probabilities, but rather a classical analysis with no restriction on measurement precision of the classical EM fields.)  Of course, the result of this analysis will disagree with actual experiments -- it will certainly not result in entangled photons -- but it is necessary to test the JORCA conjecture.  The first step is to consider the details of parametric down conversion (PDC).

\subsection{Stimulated Parametric Down Conversion}

Pairs of entangled photons are commonly generated via PDC in a non-linear medium.  As for many quantum processes, PDC has an evident classical analog where three-wave mixing transfers energy from one input EM wave into two output EM waves.  (Other, less-evidently-classical ways to generate entangled photons will be discussed in section IV.)  The physics of classical PDC can be couched in an entirely time-symmetric framework, because the medium need not be permanently altered during the three-wave mixing process.  

For simplicity, it is convenient to assume that the EM waves in question are nearly monochromatic with well-defined spatial modes -- say, $TEM_{00}$ modes.  Inside the medium, the usual $\omega$ and $\bm{k}$ phase-matching conditions are assumed to hold; $\omega_0=\omega_1+\omega_2$ and $\bm{k}_0=\bm{k}_1+\bm{k}_2$, corresponding to energy- and momentum-conservation of the corresponding photons.\footnote{This paper will not bother to distinguish the change in $\omega$ and $\bm{k}$ as the fields enter or leave the gain medium because it is not relevant to any of the calculations; both classical and quantum theory treat this change in essentially the same manner.}  (Here the $0$ subscript corresponds to the input pump beam, and $1$ and $2$ correspond to the generated lower-energy photons.)    

It is much easier to analyze classical three-wave mixing if the envelope of each wave can be described by a single complex field amplitude $E_i(t)$ (complex to encode classical phase information).  Assuming the waves can all interact as complete systems, there is no essential need to change the shape of the envelope with time or add additional spatial parameters to $E_i$.

Given these assumptions, and the lack of additional nonlinearlities, the set of equations governing three-wave mixing in PDC is simply \cite{alber1999}

\begin{eqnarray}
 \frac{dE_0}{dt} &=& i \gamma \omega_0 E_1 E_2 \nonumber \\
 \frac{dE_1}{dt} &=& i \gamma \omega_1 E_0 E_2^* \\
 \frac{dE_2}{dt} &=& i \gamma \omega_2 E_0 E_1^*.\nonumber 
 \end{eqnarray}
 
 Here $\gamma$ is a positive coupling strength characterizing the three-wave mixing.  Along with the conserved quantity corresponding to the total energy, other exactly conserved quantities are given by the Manley-Rowe relations:
 \begin{eqnarray}
 \frac{d}{dt} \left[ \frac{|E_1|^2}{\omega_1} - \frac{|E_2|^2}{\omega_2} \right] &=& 0 \nonumber \\
  \frac{d}{dt} \left[ \frac{|E_0|^2}{\omega_0} + \frac{|E_1|^2}{\omega_1} \right] &=& 0 \label{eq:threemodegain}\\
   \frac{d}{dt} \left[ \frac{|E_0|^2}{\omega_0} + \frac{|E_2|^2}{\omega_2} \right] &=& 0 \nonumber
   \end{eqnarray}
   
   The Manley-Rowe relations follow directly from the classical three-wave mixing equations without any quantization, but they can be interpreted as corresponding to a conservation of photon number, since intensity per frequency is proportional to a classical measure of photon flux.  Notice that the photon flux from the highest-frequency mode can be converted into an identical photon flux in each of the other modes; the reverse of this can also occur.  As we are interested in photons, rescaling the amplitudes to $A_i=E_i / \sqrt{\omega_i}$ is appropriate, and further simplifies the three-wave mixing equations when expressed in terms of $A_0$, $A_1$, and $A_2$.  (For the rest of this paper, the square of these scaled-amplitudes $|A|^2$ will be termed the field ``intensity'', rather than ``photon flux'', to remind the reader that it refers to a classical field quantity.)
   
   The parameter regime relevant for the generation of single photons is $A_0\gg A_1$ and $A_0\gg A_2$, even after amplification.  For all practical purposes, $A_0$ remains constant (energy, of course, is still conserved as noted above).  The phase of the pump beam is arbitrary, so $A_0$ can be defined to be real and positive.  Folding all the constants (including $A_0$) into a real constant $\alpha$, this reduces to the easily-solvable equations for the other two classical field amplitudes:
   \begin{eqnarray}
 \frac{dA_1}{dt} &=& i \alpha A_2^* \label{eq:twomodegain}\\
 \frac{dA_2}{dt} &=& i \alpha A_1^* \nonumber .
 \end{eqnarray}
 
 Classically, there is no amplification at all without initial input seed values $A_{1i}$ and/or $A_{2i}$ at $t=0$; even then, gain is not always guaranteed, depending on the relative seed phases.  In terms of these initial values, over some amplification time $T$, the final field values $A_{1f}$ and $A_{2f}$ are
    \begin{eqnarray}
A_{1f} &=&  A_{1i} \cosh(\alpha T) + i A_{2i}^* \sinh(\alpha T) \label{eq:A1} \\
A_{2f} &=&  A_{2i} \cosh(\alpha T) + i A_{1i}^* \sinh(\alpha T). \label{eq:A2}
 \end{eqnarray}
 
 The time-symmetry of this process becomes evident by inverting these equations to reveal the essentially-similar relationship
     \begin{eqnarray}
A_{1i} &=&  A_{1f} \cosh(\alpha T) - i A_{2f}^* \sinh(\alpha T) \label{eq:A1i} \\
A_{2i} &=&  A_{2f} \cosh(\alpha T) - i A_{1f}^* \sinh(\alpha T). \label{eq:A2i}
 \end{eqnarray}
 With these results, it is possible to classically analyze a typical entanglement preparation procedure, for which it is not always clear which two output modes will be amplified via PDC.  At minimum, entanglement procedures require that two different pairs of photon modes might become populated, a situation that will now be considered in detail.
 
 \subsection{Extension to four output modes}
 
 One well-known scheme to generate entangled photons uses Type II PDC, where the resulting photons are forced (by phase-matching) to have one with horizontal polarization (H) and one with vertical polarization (V), as determined by the optical axis of the crystal.  The $\bm{k}$ of these photons are different, but also correlated by phase-matching.  For a classical account of this process, each input seed considered in the previous subsection must have the same polarization and $\bm{k}$ as the output, in order to be properly amplified.
 
 The crucial piece of physics which allows for entanglement generation is that there is more than one pair of photon modes which obey the phase matching conditions.  In particular, for at least one pair of directions, it is possible for the photon in the first direction to be (H), and the second to be (V), but it is also possible for a completely different pair of photon modes to be generated via PDC -- (V) in the first direction and (H) in the second.  The classical version of this scenario is simply to allow for two additional input EM waves corresponding to this latter possibility, which can also be amplified via stimulated PDC.  If the amplitudes of these latter waves are given by $A_3$ and $A_4$, then one simply has another pair of gain equations,
 \begin{eqnarray}
A_{3f} &=&  A_{3i} \cosh(\alpha T) + i A_{4i}^* \sinh(\alpha T) \label{eq:A3} \\
A_{4f} &=&  A_{4i} \cosh(\alpha T) + i A_{3i}^* \sinh(\alpha T). \label{eq:A4} 
 \end{eqnarray}
 
By combining these with (\ref{eq:A1}) and (\ref{eq:A2}), it is clear that classical EM theory predicts that one might see gain on all four of these modes, not merely two.  The Manley-Rowe relations demand that the same intensity amplification is seen in $A_1$ and $A_2$, and also that some other equal gain is seen in both $A_3$ and $A_4$, but there is no classical restriction that only one pair of these modes could be amplified, and no restriction on the amount of amplification.  (There is no shortage of energy provided in the pump beam $A_0$, given the usual low conversion factors from the pump beam to the single output photons.)

Nonetheless, there is also no \textit{quantum} restriction that only two of these modes should be amplified.  Although a typical entanglement experiment may post-select on cases where only two photons are measured, it is certainly also possible to have four photons (one of each of these modes) produced from such a preparation.  This paper does not concern such rare-but-possible outcomes; the issue at hand is the outcomes which are impossible according to quantum theory.  And in the above example, it should be impossible to measure a single, horizontally-polarized photon emitted in both directions.  

A bit of logic reveals that the classical analog also has this very same restriction.  If both output fields are purely horizontally polarized, then the vertically polarized components must have not experienced any gain. But, due to the Manley-Rowe relations, this lack of gain in these two modes ($A_2$ and $A_3$) must also correspond to a lack of gain in both horizontally-polarized modes ($A_1$ and $A_4$).  So the only way one could get an outcome corresponding to $\ket{HH}$ is if the input seeds were also purely horizontally polarized, with the same input intensity as output intensity, for a net amplification of zero.  The JORCA conjecture says that the only possible outcomes are those with a positive total gain of energy from the pump beam $A_0$, and such a restriction would evidently rule out the $\ket{HH}$ result.  (The $\ket{VV}$ result is ruled out by identical logic.)

At this stage, this result may seem no better than the \textit{ad hoc} model from the Introduction; it is generally trivial to arrange classical anti-correlations.  But this model now predicts other anti-correlations in other measurement bases, for which a classical analog might seem to be less likely.  The central fact which prevents amplification in certain cases is that the gain indicated by Eqn. (\ref{eq:A1}) is determined by the relative amplitudes and phases of the seed waves, $A_{1i}$ and $A_{2i}$.  Certain amplitude/phase combinations necessarily lead to energy \textit{loss} to the pump beam $A_0$, and these combinations are then impossible to classical amplify.  Several examples of this scenario will now be formally analyzed.

\section{Specific Examples}

\subsection{Maximally-Entangled States}

We will begin by revisiting the Type II PDC procedure discussed immediately above, post-selected such that exactly two photons are eventually detected, one with wavevector $\bm{k}_1$ and another with wavevector $\bm{k}_2$.  (These photons will be named ``\#1'' and ``\#2'' respectively.)  The phase-matching constraints require that either \#1 is horizontally polarized and \#2 is vertically polarized -- call this state $\ket{HV}$ -- or else \#1 is vertically polarized and \#2 is horizontally polarized -- call this state $\ket{VH}$.  According to quantum theory, such a preparation procedure can therefore result in the maximally-entangled state $\ket{\psi}=(\ket{HV}+\ket{VH})/\sqrt{2}$.\footnote{This hides a phase ambiguity; there should be a $exp(i \theta)$ between the two terms, with $\theta$ determined by experimental details (generally measured empirically rather than calculated).  Similarly, the classical analog has four PDC-determined phases, one for each $A_i$ term.  Most of these phases are irrelevant to the analysis, but a certain combination relates to $\theta$.  It transpires that the analog holds when all of these phases are set to zero; this is asserted without proof for this initial example.  Even without a proof, the classical analog is sufficient to explain the case where $\theta$ is measured empirically instead of calculated.}

For maximal information at measurement, each single photon detector should not merely be placed behind a polarizing filter, but behind a polarizing \textit{beamsplitter} which separates the two polarizations onto two different paths, each leading to a single-photon detector.  The detection of a single photon therefore implies that one of these two detectors does not fire.  For the purposes of the classical analog it is reasonable to take this non-event to be a measurement of zero electromagnetic field energy in the corresponding EM mode.  (See Figure 1, and related discussion in the final section.)

\begin{figure}[htbp]
\begin{center}
\includegraphics[width=8cm]{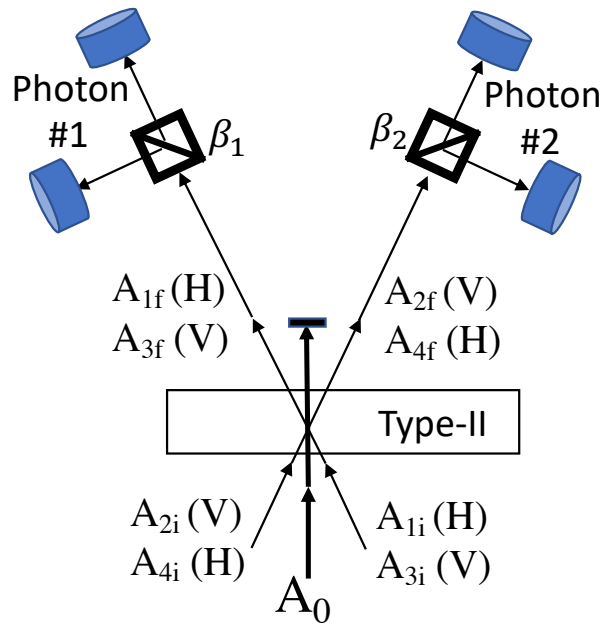}
\caption{The geometry of a simplified stimulated PDC experiment.  Four input EM waves with horizontal (H) or vertical (V) polarization are amplified by a pump beam $A_0$.  The amplified fields enter a polarizing beamsplitter, set at some angle $\beta$ to distinguish two orthogonal polarization angles.  One of the single-photon detectors behind each beamsplitter is seen to fire.}
\label{default}
\end{center}
\end{figure}

A more interesting quantum anti-correlation can be achieved by rotating the measurement basis of each photon to a diagonal polarization -- rotating each polarizing beamsplitter by 45 degrees ($\beta_1=\beta_2=\pi/4$).  Now each of the two photons are measured in the basis $\ket{+}=(\ket{H}+\ket{V})/\sqrt{2}$ and $\ket{-}=(\ket{H}-\ket{V})/\sqrt{2}$.  Rewriting the quantum state in this basis, one finds
\begin{equation}
\ket{\psi} = \frac{1}{\sqrt{2}} \ket{++} - \frac{1}{\sqrt{2}} \ket{--}.
\end{equation}
As expected, another anti-correlation is evident.  If measured in this diagonal basis, the two photons cannot have perpendicular polarizations -- they cannot be aligned either on $\ket{+-}$ or $\ket{-+}$.  This perfect anti-correlation disappears if either of the two polarizers is rotated even slightly relative to the other.

Remarkably, the classical model also reveals this same anti-correlation, although it manifests through a very different chain of logic.  As detailed in the previous section, the stimulated PDC occurs in phase-matched pairs of H and V modes.  Let $A_1$ and $A_2$ be the classical EM amplitudes which correspond to $\ket{HV}$ -- meaning that $A_1$ is the amplitude of the horizontally polarized mode along $\bm{k}_1$, and $A_2$ is vertical along $\bm{k}_2$.  Similarly, let $A_3$ and $A_4$ correspond to $\ket{VH}$.  These pairs of modes can be amplified from unknown input EM waves according to the classical amplification equations; (\ref{eq:A1}), (\ref{eq:A2}), (\ref{eq:A3}) and (\ref{eq:A4}).  The resulting EM waves corresponding to $A_{1f}$ and $A_{3f}$ evidently correspond to the measured photon \#1, and $A_{2f}$ and $A_{4f}$ correspond to the measured photon \#2.

Recall that in the classical analog, we are requiring the measured fields to be interpreted classically.  These fields are detected behind exactly one output channel of a polarizing beamsplitter; the only classical interpretation of this statement is that the fields are diagonally polarized.  Classically, diagonal polarization is a simple superposition of classical H and V waves; requiring the same intensities of these two modes, either in phase or out of phase to determine which diagonal.  (It should be of no surprise that this classical statement essentially matches the above quantum definition of $\ket{+}$ and $\ket{-}$.)  For each possible pattern of detection -- one particular diagonal for each detector -- the question is whether one might explain the zero-probability cases in a classical manner.

A close examination of the classical amplification equations reveals that certain output combinations are simply not possible -- the very same zero-probability cases of conventional quantum theory.  There are various ways to prove this, but the most elegant is to work backwards from the outcome.  Consider the hypothetical case where the detectors have revealed the $\ket{+-}$ outcome combination.   The classical interpretation of photon \#1 implies $A_{1f}=A_{3f}$ (in phase), and for \#2 this implies $A_{2f}=-A_{4f}$ (out of phase).  With these relationships, one can solve Eqns. (\ref{eq:A3}) and (\ref{eq:A4}) for the input fields:
 \begin{eqnarray}
A_{3i} &=&  A_{1f} \cosh(\alpha T) + i A_{2f}^* \sinh(\alpha T) \label{eq:A3i} \\
A_{4i} &=&  -A_{2f} \cosh(\alpha T) - i A_{1f}^* \sinh(\alpha T).\label{eq:A4i}
\end{eqnarray}

From (\ref{eq:A1i}) and (\ref{eq:A3i}) one finds
\begin{equation}
|A_{1i}|^2 + |A_{3i}|^2 = 2|A_{1f}|^2 \cosh^2(\alpha T) +2|A_{2f}|^2 \sinh^2(\alpha T).
\end{equation}
Note the cross terms cancel exactly.  Similarly, from (\ref{eq:A2i}) and (\ref{eq:A4i}),
\begin{equation}
|A_{2i}|^2 + |A_{4i}|^2 = 2|A_{2f}|^2 \cosh^2(\alpha T) + 2|A_{1f}|^2 \sinh^2(\alpha T).
\end{equation}
The sum of the left sides of these last two equations represents the total input intensity to the classical PDC, $I_{in}$ and the sum of the right sides is at least as big as the total output intensity, $I_{out}=|A_{1f}|^2+|A_{2f}|^2+|A_{3f}|^2+|A_{4f}|^2$.  Therefore this particular PDC has negative (or zero) gain, and cannot represent amplification at all.  

The conclusion is that there is no possible set of classical input seeds which will both be amplified and lead to an observed $\ket{+-}$ joint outcome.  Classical EM theory can therefore explain the absence of these outcomes, without the need for introducing entanglement or indeed postulating any nonlocal connection between the wings of the experiment.  The same conclusion holds for the $\ket{-+}$ outcome, but as in quantum theory, any slight adjustment of the polarization angle will change this result.  If $A_{1f}=exp(i\beta) A_{3f}$ and $A_{2f}=-exp(i\beta) A_{4f}$, new cross terms appear in the above calculation which allows for a net-amplification solution.

Other evident photon anti-correlations from this same state follow in a similar manner; measurements of circular and elliptical polarization are achieved by virtue of phase plates, which affect the classical analog in the appropriate way to allow the anti-correlation to persist in these other measurement bases.  Still, these measurements of maximally entangled states are all essentially similar, and these classical results might be thought to follow from a symmetry consideration rather than any deep connection between the quantum and classical experiment.  With this in mind, it is instructive to examine a curious anti-correlation observed only in partially-entangled states.

\subsection{A Partially-Entangled Scenario}

According to QM, the collapse of a partially-entangled state is often asymmetrical, meaning the two possible outcomes of photon \#1 (in some measurement basis) will each collapse photon \#2 into some state, but these two states will not generally be orthogonal.  If photon \#2 is measured in the basis of one of these possibilities (and not the other), a curious anti-correlation will result.  In this special basis, it will generally be the case that three joint outcomes will all have non-zero probabilities, but the fourth combination is strictly forbidden.  Analyzing this scenario via classical EM theory can also explain this unusual anti-correlation.

Consider the partially entangled state
\begin{equation}
\label{eq:psi'1}
\ket{\psi'} = \frac{3}{5} \ket{HV} + \frac{4}{5} \ket{VH}.
\end{equation}
Let photon \#1 be measured in the same basis as immediately above, $\ket{+}=(\ket{H}+\ket{V})/\sqrt{2}$ and $\ket{-}=(\ket{H}-\ket{V})/\sqrt{2}$.  But have photon \#2 measured in the new basis
\begin{eqnarray}
\ket{+'}&=&\frac{4}{5}\ket{H}+\frac{3}{5}\ket{V} \\
\ket{-'}&=&\frac{3}{5}\ket{H}-\frac{4}{5}\ket{V}
\end{eqnarray}
Rewriting $\ket{\psi'}$ in these bases, one finds
\begin{equation}
\ket{\psi'} = \frac{1}{\sqrt{2}}\ket{++'}-\frac{7}{25\sqrt{2}} \ket{-+'} - \frac{24}{25\sqrt{2}} \ket{--'}.
\end{equation}
The fourth possible joint outcome, $\ket{+-'}$, is predicted to have zero probability of occurring.

The physical implementation of this experiment can again be represented by Figure 1, and can again be analyzed in the context of classical EM theory.  Only two changes are needed from the previous example.  First, the two pairs of phase-matched PDC modes cannot be amplified to the same value -- looking at Eqn. (\ref{eq:psi'1}), one can see that for any classical analog $A_3$ and $A_4$ must be amplified more than $A_1$ and $A_2$.  This is easily accomplished; indeed, it is generally unlikely that the two pairs of phase-matched modes would have precisely the same gain coefficient $\alpha$.  The solution is to set $\alpha T\to 0.6\epsilon$ in the equations relating $A_1$ and $A_2$, and $\alpha T\to 0.8\epsilon$ in the equations relating $A_3$ and $A_4$.  (Here $\epsilon$ is an arbitrary unitless gain parameter.)  In the small gain limit, this gives the appropriate gain ratio for the EM field amplitudes.

The other simple change is to rotate the polarizing beamsplitter for photon \#2 such that the detectors are sensitive to the $\ket{+'}, \ket{-'}$ basis.  This is accomplished by setting $cos(\beta_2)=4/5$ instead of the previous value of $\beta_2=\pi/4$ (see Figure 1).  With this setting, given the measured outcome of photon \#2 corresponding to $\ket{-'}$, one can back-evolve the implied classical EM field to reveal that $0.6 A_{2f} = -0.8 A_{4f}$.  So a joint measurement of $\ket{+-'}$ implies both this relationship, as well as the same $A_{1f}=A_{3f}$ from the previous example.

Following the same procedure as before (except using the different PDC gains of this new example), the total input intensity can be calculated from Eqns. (\ref{eq:A1i}), (\ref{eq:A2i}), (\ref{eq:A3i})and (\ref{eq:A4i}), along with the $\cosh(2x)$ and $\sinh(2x)$ identities:
\begin{eqnarray}
\label{eq:PEI}
I_{in} &=&  (|A_{1f}|^2+|A_{2f}|^2) \cosh(1.2\epsilon) + (|A_{3f}|^2+|A_{4f}|^2) \cosh(1.6\epsilon)   \\ 
& & - Im(A_{1f}A_{2f})[\sinh(1.2\epsilon) - 0.75 \sinh(1.6\epsilon)]. \nonumber
\end{eqnarray}
The first two terms on the right clearly reduce to the total output intensity $I_{out}$ as $\epsilon\to 0$, with a positive $\epsilon^2$ correction in the small gain limit.  One might think that the final term could reduce the total right hand side to a value less than $I_{out}$, as it could be negative, and those $\sinh$ terms scale like $\epsilon$ instead of $\epsilon^2$.  However, the precise numerical values involved exactly cancel the $\epsilon$ scaling in the final term, leaving $\epsilon^3$ as the dominant factor in the small gain limit.  Therefore, as before, $I_{in}>I_{out}$, and no solution with a net amplification is possible.

Notice that this cancellation requires the measurement basis to ``match'' the unequal gains from the classical PDC, or else this precise cancellation in the $\epsilon$ terms would not occur, and there would be a solution with $I_{in}<I_{out}$.  Even in the high-gain limit, miniminizing the right-hand side of (\ref{eq:PEI}) requires $3|A_{2f}| = 4|A_{1f}|$ (the expected intensity ratio), but this still cannot reduce the total to less than $I_{out}$.  The conclusion is that for the particular outcome $\ket{+-'}$, there is no possible set of classical EM input fields which could be amplified by the PDC process, matching the quantum conclusion that this joint outcome should be strictly forbidden.  The other three outcomes all remain possible; the probabilities of these outcomes will be discussed in a follow-up paper, as they are not the anti-correlations discussed here.

\subsection{Hardy's Nonlocality Proof } 

One interesting application of zero-probability outcomes is employed in Hardy's proof of nonlocality without inequalities \cite{hardy1993}.  We have already shown that for the state $\ket{\psi'}$, if measured in a certain pair of bases, the joint measurements corresponding to $\ket{+-'}$ have probability zero.  The structure of Hardy's proof requires there are two other probability-zero outcomes if the basis of only one of these measurements is changed in a particular way.  Specifically, the probability of the outcomes corresponding to both $\ket{\chi +'}$ and $\ket{-\phi}$ should be zero, for some states $\ket{\chi}$ and $\ket{\phi}$.  For the state $\ket{\psi'}$, it so happens that the special basis for the first of these cases includes
\begin{equation}
\label{eq:337}
\ket{\chi}=\frac{16}{\sqrt{337}}\ket{H}-\frac{9}{\sqrt{337}}\ket{V} ,
\end{equation}
measured on the first photon.  The special basis for the second of these cases includes
\begin{equation}
\ket{\phi}=\frac{3}{5}\ket{H}+\frac{4}{5}\ket{V} ,
\end{equation}
measured on the second photon.

For the classical analog to retain the essentials of Hardy's proof, the joint probability of both $\ket{\chi +'}$ and $\ket{-\phi}$ must be zero.  The second case is quite trivial, given the previous subsection.  A measurement of the classical polarization corresponding to $\ket{-}$ in the first photon corresponds to the condition $A_{1f}=-A_{3f}$, and a polarization corresponding to $\ket{\phi}$ in the second photon corresponds to $0.6 A_{2f} = 0.8 A_{4f}$.  But this is just two reversed signs from the previous analysis, and the structure of Eqn. (\ref{eq:PEI}) continues to rule out this possibility.  

The other zero-probability case (\ref{eq:337}) appears quite different, as a polarization corresponding to $\ket{\chi}$ in the first photon implies $9 A_{1f}=-16A_{3f}$ and a polarization corresponding to $\ket{+'}$ in the second photon implies $0.8 A_{2f} = 0.6 A_{4f}$.  But a closer examination reveals that the numerical factor in the final term of Eqn. (\ref{eq:PEI}) is determined by the ratio $A_{3f}A_{4f}/(A_{1f}A_{2f}) = -0.75$.  This ratio is precisely the same for this third case, so that the calculation of $I_{in}$ yields the same Eqn. (\ref{eq:PEI}) despite the superficially different values.  

A generalization of these results -- beyond just this particular numerical example -- is presented in the next section.  But given that these results follow from classical electromagnetism, it is worth explaining why Hardy's proof of nonlocality does not extend to classical physics. The crucial difference is that in classical physics, EM fields are not absorbed in discrete quantities and this prevents Hardy's proof from going through. In particular, the central argument in Hardy's proof uses the point that for a given measurement basis, with an impossible outcome such as $\ket{+-'}$, a measurement of $\ket{+}$ on the first photon necessarily implies a measurement of $\ket{+'}$ on the second photon.  (Conversely, a measurement of $\ket{-'}$ on the second photon implies a $\ket{-}$ result on the first.)  But while these implications follow from quantum theory, where only a certain number of outcomes are known to be allowed, they do not follow from CEM alone, as  classical theory has no such discrete restriction on the outcomes.  For example, there is no classical reason why half a photon's worth of EM energy could not be detected on each of four final detectors.  Therefore the probabilities assigned by CEM to \emph{non-zero} probability outcomes do not always match the probabilities assigned by quantum mechanics, and thus while this analysis recovers the zero-probability cases utilized by Hardy's proof, the rest of the argument can't be made and so there is no contradiction between Hardy's proof and the fact that the CEM model is local. 
 
\section{Generalizations}

\subsection{Proof for general qubit states \label{proof1}} 

In fact, it is possible to show that for any initial state of the form $| \psi \rangle = a | H V \rangle + b | H V \rangle$ for real numbers $a, b$ satisfying $a^2 + b^2 = 1$, and any pair of outcomes of the form $ | \phi \rangle = c |H \rangle + d | V \rangle, | \chi \rangle =  f |H \rangle + g | V \rangle$, for complex numbers $c, d, f, g$ satisfying $ |c|^2 + |d|^2 = 1, |f|^2 + |g|^2 = 1$, the classical model will predict zero probability whenever quantum mechanics predicts zero probability. 

First observe that quantum-mechanically, the probability to obtain outcome  $\phi$ to a measurement on the first photon and $\chi$ to a measurement on the second photon, when performing measurements in bases which respectively include $\ket{\phi}$ and $\ket{\chi}$, is the square of $(\langle \psi ( | \phi \rangle \otimes | \chi |\rangle) = a^* c  g  + b^* d  f = a c  g  + b d  f$.  Evidently, this probability is zero iff $acg=-bdf$.

Now, turn back to the classical model.  The state $a | H V \rangle +  b | H V \rangle$ is represented in the classical model by setting $\alpha T = a \epsilon$ for the $A_1$ and $A_2$ modes, and setting $\alpha T =  b \epsilon $ for the $A_3$ and $A_4$ modes.  As before, one can calculate the total input intensity on all four modes $I_{in}$ in terms of the output intensities using Eqns. (\ref{eq:A1}), (\ref{eq:A2}), (\ref{eq:A3}), (\ref{eq:A4}).  In general, this relationship is 

\begin{eqnarray}
I_{in} &=&  (|A_{1f}|^2+|A_{2f}|^2) \cosh(2a\epsilon) + (|A_{3f}|^2+|A_{4f}|^2)  \cosh(2b\epsilon)  \\ 
& &  - Im(A_{1f} A_{2f}) \sinh(2a \epsilon) - Im(A_{3f} A_{4f}) \sinh( 2b \epsilon) . \nonumber
\end{eqnarray}

To first order in $\epsilon$, the $\cosh$ terms go to unity, and the total gain in the PDC is approximately

\begin{equation}
    I_{out}-I_{in} \approx  2\epsilon [a \, Im(A_{1f} A_{2f}) + b \, Im(A_{3f} A_{4f}) ].
\end{equation}

It is this quantity which can never be positive for the outcomes corresponding to zero probabilities.  The outcome $\phi$ corresponds to final fields satisfying $d A_{1f} = c A_{3f}$ and the outcome $\chi$ corresponds to final fields satisfying $f A_{2f} = g A_{4f}$. (Measuring complex coefficients simply corresponds to an arrangement with a phase plate before the polarizing beamsplitter; see \cite{tyagi2022} for details.)  For these outcomes, 

\begin{equation}
I_{out}-I_{in} \approx 2\epsilon \left[a Im(A_{1f} A_{2f}) + b Im\left( \frac{df}{cg} A_{1f} A_{2f} \right) \right] .
\end{equation}

If the quantum probability for this outcome is zero, we have $acg = -dfb$ and hence $(df)/(cg)=-a/b$, setting the gain to zero.  So in this small-$\epsilon$ limit, there are no values of the input fields which lead to a positive gain for this set of outcomes, meaning that the classical model also predicts probability zero for this case.    

Further generalizations are possible.  By adding a phase plate immediately after the PDC, imposing a net phase delay of some angle $\delta$ on the $A_3$ mode (only), one would transform the corresponding quantum state to the generic Schmidt-basis state
\begin{equation}
\label{eq:schmidt}
    a | H V \rangle +  b e^{i\delta} | H V \rangle.
\end{equation}
This would change the quantum probabilities; they now would only be expected to go to zero if $acg = -dfb \exp(-i\delta)$.  But this could also be tested by measuring the first photon in a different basis, where $d\to d\exp(-i\delta)$.  Such a measurement, in turn, would correspond to adding a phase plate delay of an angle $-\delta$ on the $A_3$ mode before it encountered the polarizaing beamsplitter.  But evidently, the classical effect of this extra phase plate would exactly cancel the original phase plate, and so again all of the zero-probability predictions would remain correct.

We observe that via the Schmidt decomposition, any pure two-qubit state can be written in the form of Eqn. (\ref{eq:schmidt}) for some choice of orthogonal states.  By rotating one's coordinate system (see \cite{tyagi2022} for details), any pure two-qubit state can be analyzed in this same manner.  Furthermore, every mixed state can be written as a convex decomposition of a set of pure states, so we can also produce all mixed states by means of probabilistic procedures (selecting the preparation procedure according to some probability distribution, etc.). 

\subsection{Other methods of creating entanglement} 

We have focused so far on entangled photons generated via parametric down-conversion. However, there exist other ways of creating entangled states, and given the striking similarity in the behavior of entangled systems in different media, one would naturally like to offer accounts of other forms of entanglement which are suitably analogous to the account offered for parametric down-conversion. 

The entangled states used in the early Aspect experiments\cite{PhysRevLett.49.1804} (and earlier experiments including those of Kochens and Commins\cite{kocher1967polarization} and Freedman and Clauser\cite{Freedman1972ExperimentalTO}) were created using atomic cascades in calcium.  Consider for instance a $J\!=\!0 \rightarrow J\!=\!1 \rightarrow J\!=\!0$ atomic cascade.  The spontaneous decay of an excited atom emits two photons, and when the photons are emitted in exactly opposite directions they must have orthogonal polarization, conserving both linear and angular momentum.  These photons have been measured to act as entangled states.  

As in the case of parametric down-conversion, these experiments can be given a classical analog based on the amplification of seed waves. The simplest classical analog employs the ``electron oscillator model'' (see Cray \textit{et al.} \cite{cray1982}, and references therein) to represent the atom as charged masses on springs, in a localized oscillation which can be regarded as a classical wave with zero wavevector, $\bm{k}_0 = 0$.  Cray  \textit{et al.} assume that there are incoming EM ``seed'' waves which are amplified during this process - the gain is the result of energy being extracted from the charged oscillator  by means of interference between the incoming waves and the field.\cite{cray1982} Cray \textit{et al.} take it that the incoming waves are known, so they regard this process as a classical analog of stimulated emission, but if we assume that the seed waves are unknown then we can instead take it as a classical analog of \emph{spontaneous} emission, thus reproducing the qualitative features of  an atomic cascade.  By enforcing the $J\!=\!0\to J\!=\!0$ cascade, there is no net momentum change in the charged oscillator, so the energy must be emitted in pairs of modes with equal and opposite momentum, i.e. we have $\bm{k}_1 + \bm{k}_2 = \bm{k}_0 = 0$.

While this physical scenario may seem quite different from the PDC described earlier, at the level of the gain equations this is effectively just another example of three-wave mixing, with the zero-wavenumber atomic oscillator effectively serving as the ``pump beam''.  As in the PDC case, energy conservation (corresponding to the Manley-Rowe relations) implies that both members of the pair experience the same gain, which means that it's impossible to produce the outcomes $\ket{HH}$ or $\ket{VV}$ with net gain greater than zero. If we demand that the only possible outcomes are those with a positive total gain of energy from the atomic pump, it follows that only $\ket{HV}$ and $\ket{VH}$ outcomes are possible in this basis. 

There is one slight difference between this example and the PDC case, as can be seen if one measures in the $| + \rangle, | - \rangle$ basis.  Due to the condition on orthogonal polarizations, it's now impossible to obtain the outcomes $\ket{++}$ or $\ket{--}$ with net gain greater than zero, rather than the previous case where it was impossible to obtain either $\ket{+-}$ or $\ket{-+}$.  But this is simply due to the fact that this cascade preparation corresponds to $\ket{HV}-\ket{VH}$ rather than $\ket{HV}+\ket{VH}$, and these two cases can be transformed into each other with a simple $\pi$ phase delay on one of the $\ket{H}$ or $\ket{V}$ modes.  With this adjustment, the amplified modes obey equations precisely analogous to (\ref{eq:twomodegain}), and all of the above results go through. 

In atomic cascades we typically have large ensembles of excited atoms, but modern techniques have made it possible to create entanglement using sources which produce only one pair of photons at a time. For example, semiconductor quantum dots employ a radiative cascade as in the Aspect case, but the dots are constructed such that only one electron is excited at a time, and therefore (quantum-mechanically) exactly one pair of photons is created when the electron decays. However, a classical analog of these scenarios would still utilize an oscillating classical charge to represent the electron, with no quantization restriction on the output fields.  In this description, then, multiple EM modes could be amplified by a single oscillator, just as in the classical analog of PDC.  

Another class of entanglement schemes starts with two independent photons, and then entangles them via a two-qubit ``gate'' such as a CNOT.  At first glance this would seem to be quite different from a three-wave mixing scenario, but a closer examination reveals that many such gates are implemented by using another pair of entangled photons \cite{pittman2001} -- often generated through a three-wave mixing process such as PDC.  Applying the above classical procedure to this other pair of photons therefore restricts the interactions which are allowed to occur in the gate, and can in turn lead to a classical account of any anti-correlations observed in the gate's outputs.

Nevertheless, there is one category of photon correlations for which the above classical analogs do not apply: the Hong-Ou-Mandel (HOM) effect \cite{hong1987}, and other related phenomena which rely on the fundamental indistinguishability of certain photons.  This effect is utilized in various ``linear'' entanglement schemes \cite{obrien2003}, but it disappears as the photons begin to differentiate in either frequency or temporal overlap.  As the above classical discussion did not assign any importance to these parameters, it would be impossible for it to capture the HOM effect without introducing a measure of indistinguishability into the classical analysis.  It is an open question whether a classical analog might be extended to such HOM-based entanglement generation procedures.

\section{Analysis}

The above results indicate strong support for the original JORCA conjecture, that joint outcomes require classical analogs.  Specifically, for any pair of photons generated by a three-wave-mixing process with a net gain, the perfect anti-correlations exhibited by quantum theory are also implied by classical EM.  The ability of classical EM theory to recover a result that seems like it should require a non-local QM calculation is surprising enough to warrant some further analysis.  But first, it is worthwhile to note some of the formal differences between QM and EM.

Classical EM theory is covariant, capable of describing the same events with the same equations under arbitrary Lorentz boosts.  On the other hand, QM does not work in this manner; the order in which space-like separated measurements occur affects the intermediate steps in the calculations of entanglement phenomena (even if the end result always happens to end up the same).  For example, in the partially-entangled state examined in Section III, the computational account of \textit{why} the $\ket{+-'}$ state has a zero probability looks quite different depending on which photon is measured ``first''.  This is in contrast with the above classical analog, which is just as covariant as classical EM itself.

An even more pronounced difference between EM and QM concerns the sorts of mathematical objects which are presumed to correspond to physical quantities.  Classical EM has an ontology of local beables; the fields $\bm{E}(x,y,z,t)$ and $\bm{B}(x,y,z,t)$ appear to describe localized events, and thus most physicists are happy to take the mathematics literally and  imagine these fields `existing' at the corresponding location in spacetime, matching the notion of a pointlike event in relativity.  In classical EM, one can talk about quantities distributed over space and time, such as a classical outcome corresponding to the measurement $\ket{+-}$, but this is just a shorthand for two distinct events:  a classical field configuration corresponding to $\ket{+}$ in one place, and another classical field configuration corresponding to $\ket{-}$ somewhere else.

On the other hand, the mathematics of QM cannot be reduced to local beables:  for an entangled state, the full wavefunction $\ket{\psi}$ cannot be reduced to a classical conjunction of events.  So if we take this mathematics literally, we may be inclined to conclude that reality consists of unseparable, spatially-non-local ``states''.  (This is far from the only interpretation of the QM mathematics; for example there is a sizable literature on how quantum states are more naturally viewed as states of knowledge \cite{spekkens2007,leifer2013,wharton2014}.) Moreover, it is common to imagine that the nonlocal behaviour exhibited in Bell experiments is in some way a consequence of the existence of these unseparable, non-local states. From this point of view, it may seem surprising that classical EM theory using only local beables could possibly explain the same anti-correlations as QM.  After all, classical EM has no direct connection between the two wings of the experiment.  A potential resolution to this apparent mismatch will be addressed next.  

\subsection{Comparison with the Path Integral}

Although conventional state-based QM has profound differences with classical EM (as outlined immediately above), the path integral offers an alternative  approach to QM which looks much closer to the covariant classical theory.  The path integral
is as covariant as the classical action on which it is based, and also does not rely on ``states'' in any way.  Applying the path integral analysis to entanglement scenarios can be used to calculate two-photon correlations without relying on any direct connection between spacelike-separated events\cite{sinha1991,tyagi2022}  - indeed, all the ``histories'' used in such approaches consist of worldlines that track the very same classical trajectories utilized in the classical EM analog.  Such a connection between the classical analog and the path integral makes it seem less surprising that both models can recover the same zero-probability results.

But there are still notable mismatches between classical EM theory and path integrals.  The former is often assumed to be a computation of ``one real history'' -- one actual collection of spacetime-localized fields.  The mathematics of the path integral, on the other hand, assigns a complex amplitude $\mathcal{E}_j$ to each of several different histories (ending at the same outcome $z$), and then sums and squares the amplitudes to calculate the relative probability of that outcome, $P(z) \propto | \sum \mathcal{E}_j |^2$.  Another difference is that the path integral treats photons as discrete particles, forcing them along a single path at each potential branching point.  In classical EM, on the other hand, the fields will naturally divide and/or recombine at a beamsplitter.  Somehow, despite these differences, both approaches seem to agree on the above anti-correlation cases where $P(z)=0$.

A closer examination reveals that, to some extent, these differences cancel each other out.  Each particle trajectory used in the path integral cannot interfere with itself, but interference terms appear between different histories when computing $P(z)$.  Classical EM does not have this feature, but it is not starting with particle trajectories in the first place; the field history contains interference simply because it represents the physics of a single classical field.  Even though these two types of interference appear in different contexts, they still refer to the same spacetime locations in the same experimental geometry, so despite the differences there is a structural commonality between the two approaches which explains why they recover the  same perfect cancellation in certain circumstances. 

For example, consider the maximally-entangled example from section 3.1, with both photons measured in the $\ket{+},\ket{-}$ basis.  The path integral account of this experiment \cite{sinha1991,tyagi2022} combines the amplitude from two equal-amplitude histories; first $\mathcal{E}_{12}$, where photons are emitted along $A_1$ and $A_2$, and second $\mathcal{E}_{34}$, where photons are emitted along $A_3$ and $A_4$.  (For each history, the two photons end up at the two particular detectors for which one wants to calculate the joint probability.)  Given that $|\mathcal{E}_{12}|=|\mathcal{E}_{34}|$, the probability $P(z)$ is entirely determined by the phase difference between the two amplitudes; a zero-probability results when the amplitudes are out of phase. 

It transpires that the classical model is able to  reproduce this sensitivity to the phase difference of these two particle-histories.  For measurements on the diagonally-polarized $\ket{+},\ket{-}$ basis considered in Section IIIA, it turns out that all four of the classical modes must have the same amplitude, again leaving the calculation entirely dependent on phase.  Now the key physics is encoded in the details of the classical stimulated PDC; there is a particular phase between each pair of input seeds which will lead to maximum gain, and the opposite phase will lead to maximum loss.  But a polarization measurement is also a measurement of relative phase between the pair of output modes at a particular detector.  The constraint that the total gain must not be zero therefore relates all of these phases in a way that is ultimately equivalent to the phase-dependence that appears in the path integral.

The existence of a mapping between the classical and quantum descriptions in the zero-probability case is interesting because it is notoriously difficult to  interpret the path integral in any realistic manner \cite{Wharton2016}, in part because of the way it combines completely different histories. The results of this paper demonstrate that the zero-probability cases don't in fact have to combine different histories: the same results can be recovered using just a single classical history. Thus this result represents a promising step towards  a more realistic understanding of the path integral. 

\subsection{Other Potential Implications}

The classical perspective utilized here might also inform other questions in quantum foundations, most notably the quantum-classical divide.  Consider, for instance, the way that uncertainty comes into the classical model, by taking seriously the possibility space of all unknown classical inputs.  The classical model is fully deterministic, in the usual sense that the initial conditions exactly determine the final output. However, the initial seed values ($A_{1i}$, etc.) are associated with fields too small to be detectable, so it is not \textit{computationally} deterministic:  since the exact values of these quantities aren't known, they could give rise to what appears to be random variation. 

In contrast, conventional quantum theory takes the initial state to be (in principle) completely determined by the preparation, followed by deterministic unitary evolution, such that all the uncertainty only comes in when considering measurements.  To the extent that this conventional viewpoint runs into difficulties (say, the famous ``measurement problem''), the classical perspective might seem to provide hints to alternate approaches.  Indeed, much of the analysis of ``quantum noise'' \cite{clerk2010} straddles these two perspectives, commonly discussing input vacuum states as if they were a source of random noise.  However, rigorously, there is only one QM vacuum state, not a wide array of potential inputs as in the classical case.  This classical analog might provide further motivation for reformulating quantum theory in a manner that allowed for unknowable inputs.

The framework presented here demonstrates that at least in the two-photon case, the distribution of zero-probability outcomes is not a characteristically quantum phenomenon. This has implications for ongoing discussions over Hardy's paradox, as discussed in Section III.C. It is common to interpret Hardy's result as demonstrating that quantum-mechanical nonlocality can be demonstrated using purely the distribution of zero-probability results, but since we have shown that the zero-probability cases in Hardy's case can in fact be reproduced within classical electromagnetism, it is now clear that what makes Hardy's example characteristically quantum is not the assignation of zero probabilities \textit{per se}, but rather the discrete restriction on the set of possible outcomes.  This is an intriguing indication that \emph{discreteness} may have a closer relationship with quantum no-go theorems than much of the literature would seem to suggest.

In fact, the classical perspective offers an alternative take on discreteness.  Typically, discrete quantum phenomena are taken to be ``non-classical''.  But if one allows for unknown classical inputs, the above results imply that all non-zero probability outcomes (even discrete ones) do in fact have a particular classical input which would lead to such a result.  Since the space of such classical scenarios (with unknown inputs) is much larger than the space of discrete quantum scenarios, this motivates the perspective that the vast majority of classical cases are somehow ``non-quantum'' -- or in other words, that some quantum discreteness condition constrains the classical possibility space.  This is along the lines of how Bohr-Sommerfeld quantization was utilized in the ``old'' quantum theory, but not how one typically considers discreteness in modern QM.

Finding an analog of quantization for the classical EM model is one of the main considerations that must be addressed if some classical analog might be developed for the \textit{non-zero} probability outcomes.  Indeed, developing such an analog is probably the most obvious research direction raised by the JORCA conjecture.  We will therefore end this paper with short discussion of the challenges and prospects of such a future development.  

\subsection{Open Questions and Future Directions \label{open}}

While it is not obviously impossible to modify the above classical model to account for the outcomes which are actually seen to occur in entanglement experiments, a number of serious challenges would have to be simultaneously addressed.  Clearly, any two-photon measurement of the sort analyzed here only has at the most four possible outcomes.  But classical EM has an infinite continuum of possibilities, if one allows for hidden input seeds.  So there will certainly not be a full map between histories which are allowed by classical EM and measurements which are allowed by quantum theory:  for any particular apparatus and measurement settings, only a very few classical EM histories have any hint of an analog to the quantum outcomes.

Furthermore, even a brief analysis will reveal that these few classical histories are sensitive to the precise measurement settings -- in this case, the angles of the two polarizing beamsplitters.  Changing these angles (the measurement basis) would dramatically alter the analogous classical EM histories which matched the allowed quantum outcomes. This would necessarily extend back to times before the measurement basis was chosen, including to the hidden input seeds $A_i$.  If one would like to have some eventual model in which these analogous classical EM histories were somehow special, that model would necessarily be ``future-input dependent'' \cite{wharton2020}.  In other words, the allowed values of the past input seeds would literally have to depend on future measurement settings.  Any classical model of this sort would presumably have to be action-based, able to analyze entire classical field histories ``all at once'', such as the direct action concept of Wheeler and Feynman \cite{wheeler1949}.\footnote{Such models need not permit signalling into the past, because the unknown input seeds are hidden, and would not be knowable in advance of the eventual measurement outcome.}
 
There are other technical challenges awaiting any effort to literally interpret these classical EM analogs.  In real-life PDC situations, the classical gain into any single mode is typically quite minimal -- indeed, it is this low-gain which allows for single-photon generation.  If one takes this small gain at face value, there is essentially no classical way for a full photon's worth of energy to be produced into a single mode.  (Technically any energy amplification would be possible, given a large enough input seed, but these inputs seeds are unobservable.  So any reasonable argument would restrict the unknown seeds to the quantum zero-point field regime, where they would be too weak to generate a full photon's energy in the required output mode.)

Another problem concerns some asymmetrical assumptions.  The classical analogs considered above \textit{require} unknown input fields, on top of any known fields.  But when it came to the eventual measurement, the model in this paper assumed that the measurement was perfectly informative.  (A measurement of a horizontally-polarized photon was assumed to be a measurement of an EM field with no vertically-polarized component at all.)  But if hidden fields are logically possible upon preparation, then by symmetry they should also be logically possible upon measurement.   

There are also some simplifying assumptions made in this paper that would need to be lifted in order to consider more complex cases. In particular, the proof of the result in section \ref{proof1} did assume that each classical EM wave could be parameterized by a single complex amplitude, varying with time, rather than a more realistic spatially-dependent field with a finite bandwidth.  It remains to be shown whether or not the above results will go through for such a detailed description of the classical fields. This extension to a more sophisticated field model would be necessary to search for classical analogs to the HOM effect and associated ways to produce entangled photons using only linear optics.  Such an effect sensitively depends on the precise indistinguishability of two photons, a concept which would have to be carefully translated into a classical analog.  (Without \textit{perfect} indistinguishability, the anti-correlations from such sources would not themselves be perfect.)  It also remains to be seen whether or not such phenomena would follow from this classical analysis.

Despite these unresolved issues, we still see room for future developments along these precise lines.  Indeed, were anyone to devise a model which correctly accounted for the joint probabilities of an entanglement experiment, it would have to violate at least one of the assumptions of Bell's Theorem, such as the assumption of no-future-input-dependence \cite{wharton2020}.  Being forced into a future-input-dependent model would therefore possibly be seen as an advantage for some eventual model based on classical EM, aiming to provide a more-classical account of what is happening in an entanglement experiment between preparation and measurement, without requiring any direct connection between the distant wings of the experiment.  What can be said for certain, however, is that some surprising quantum correlations have been hiding in classical EM all along -- the zero-probability joint measurements of anti-correlated photons.

\begin{acknowledgments}
The authors thank R. Sutherland and N. Argaman for useful suggestions.
\end{acknowledgments}

\bibliography{Kiyaeva}

\end{document}